\documentclass{article}
\usepackage{graphicx} % Required for inserting images

\usepackage{enumitem}
  \usepackage{natbib}
\date{\today} 
\setcitestyle{round}

\title{What Kind of Relationality does Quantum Mechanics Exhibit?} 
\date{\today}

 \author{Emily Adlam \thanks{Philosophy Department and Institute for Quantum Studies, Chapman University, Orange, CA92866, USA \texttt{eadlam90@gmail.com} }}

\begin{document}

\maketitle

\begin{abstract}

In this article I elaborate on the approach to  relational quantum mechanics suggested by  Adlam and Rovelli (2023). I suggest that this approach fills an important gap in the spectrum of relational approaches, because it posits that the relational aspects of quantum mechanics are both inherent and dynamical. I  compare this approach  to Orthodox RQM, arguing that it has a number of advantages, and I show how  some possible objections  can be resolved.
    
\end{abstract}

\vspace{3mm}

\section{Introduction \label{intro}}

Relational Quantum Mechanics (RQM) is an interpretation of quantum mechanics based on the idea that the quantum descriptions must be relativized to physical systems.  \cite{https://doi.org/10.48550/arxiv.2203.13342} have recently proposed a possible way of altering relational quantum mechanics to allow observers to share facts with one another. This proposal amounts to an alternative version of relational quantum mechanics which diverges from some previous presentations. In this article I will elaborate  on what such an alternative approach might look like, and how it can be motivated.

The standard formulation of RQM, which I will refer to as `Orthodox RQM,'  involves  a radical   form of relationalism in which all physical facts must be relativized to an observer.  By contrast, the proposal of Adlam and Rovelli, which I will refer to as `RQM+CPL,' suggests a more moderate form of relationalism. In this picture there exist absolute facts about the relations between observations made by different observers, but nonetheless quantum states are relativized, because the outcome of a quantum measurement typically depends on facts about the observer performing the measurement as well as facts about the  system being measured. That is, in RQM+CPL the relativization applies primarily to  \emph{dynamics} rather than to facts.

 My goal in this article is to show that RQM+CPL need not be understood as merely an ad hoc response to a problem in Orthodox RQM: it can be given independent motivation as a plausible interpretation of quantum mechanics in its own right. In section \ref{motivation} I will set out the general argument for relational approaches to quantum mechanics, and I will argue that there is an important gap in the spectrum of relational approaches, to be filled by an approach in which the relativization in quantum mechanics is inherent but nonetheless dynamical.  In section \ref{orthodoxCPL} I will show that RQM+CPL can be understood as a form of inherent-dynamical relationalism, and explore what that means for its ontology.   In section \ref{compare} I will compare RQM+CPL  to Orthodox RQM, arguing that it has a number of important advantages. Finally, in section \ref{objection} I will consider a few possible objections to RQM+CPL and show how they can be resolved.

\section{Relational Approaches to Quantum Mechanics\label{motivation}}

  The general argument for relationalism in quantum 
 mechanics is based on the  Wigner's Friend scenario \citep{Wigner}. Imagine  some observer, Alice, performing a measurement in the basis $\{ |0 \rangle \langle 0 |, |1 \rangle \langle 1  | \}$ on a two-dimensional system $S$  originally prepared in the state $a | 0 \rangle + b |1 \rangle$, with both $a$ and $b$ non-zero. Presumably Alice will see an  outcome to her measurement,  and the standard textbook formulation of quantum mechanics prescribes that the system is now in the state corresponding to that outcome - that is, if Alice sees the outcome $0$, then the state of system $S$ is now $| 0 \rangle$, and if Alice sees the outcome $1$ then the  state of system $S$ is now $| 1 \rangle$. 

However, if Bob observes from the outside and describes this whole interaction using unitary quantum mechanics, he will conclude that  $S$ and Alice are now in the following entangled superposition state: 

\[ \psi_{SA} = a | 0 \rangle_S | 0 \rangle_A + b |1 \rangle_S | 1 \rangle_A \]

Here, $|0\rangle_A$ and $|1 \rangle_A$ are states of Alice corresponding to `seeing outcome 0' and `seeing outcome 1' respectively. So this scenario appears to present us with a contradiction: common sense suggests that Alice has  experienced a single outcome, and yet the state assigned by Bob appears to suggest that Alice is  somehow in a mixture of experiencing two different outcomes. 

Now, one might initially hope that this contradiction may just result from different  choices of representational conventions. However this cannot be the case,  because it can be shown that Alice and Bob will now make different predictions for the outcomes of certain measurements, so it appears that we have come up against  a real empirical conflict. For example, suppose   Alice obtained outcome $0$ for her measurement - in that case  she now assigns the pure state $\psi_A = |0 \rangle_S$ to $S$, thus predicting that a measurement performed on $S$ in the basis $\{ |0 \rangle \langle 0 |, |1 \rangle \langle 1  | \}$ is guaranteed to obtain the outcome $0$. Whereas 
 Bob, regardless of Alice's outcome, will always assign to $S$ the mixed state $\psi_B = |a|^2 |0 \rangle \langle 0|_S + |b|^2 |1 \rangle \langle 1 |_S$, thus predicting that a measurement performed on $S$ in the basis  $\{ |0 \rangle \langle 0 |, |1 \rangle \langle 1  | \}$ is \emph{not} guaranteed to obtain the outcome $0$. That is, the different states $\psi_A$ and $\psi_B$ encode different predictions about the outcomes of measurements on $S$, even for the very same measurement performed  at the very same time.  

There are various options for responding to this dilemma, but in this article I will consider only responses which  maintain the first-person universality of quantum mechanics. By this I mean that they do not alter any of the predictions of unitary quantum mechanics as applied by any individual observer - in particular,  no observer will ever see any evidence of wavefunction collapse in any scenario except a measurement they have personally performed. If first-person universality holds, then Alice's assignation of the state $\psi_A$ must give the right predictions for \emph{her} measurements on $S$, and Bob's assignation of the state $\psi_B$ must give the right predictions for \emph{his} measurements on $S$ - that is, Alice and Bob  must both be  correct in using the states $\psi_A$ and $\psi_B$ respectively to make predictions for their own measurements, even though they are thereby assigning incompatible states to the very same system at the very same time. Thus perhaps the central moral of the Wigner's Friend scenario is that if we want to maintain the first-person universality of unitary quantum mechanics, we have no choice but to accept that at least at some effective empirical level, quantum states are relativized to observers.

Obviously, such relativization appears prominently in RQM and various perspectival approaches. But any other view which maintains first-person universality, such as Everett or de Broglie-Bohm, must also exhibit some form of relativization. For example,   the Everett picture has  `relative states'  associated with  branches of the wavefunction, distinct from the underlying non-relativized global quantum state \citep{Everett}. And  in the Bohmian picture we use  `conditional wavefunctions' relativized to Bohmian particles to explain the appearance of wavefunction collapse in certain scenarios \citep{D_rr_1992, articlenorsen}. In these contexts quantum states are relativized in the sense that  it is, at least at an empirical level, appropriate for different observers to use different  quantum states to predict the outcomes of \emph{their own} measurements, although things will go wrong if they start trying to use these quantum states to predict the outcomes of someone else's measurements. Thus although the Everett and de Broglie-Bohm approaches do maintain that there is always an underlying absolute quantum state, they also sometimes include effective relational quantum descriptions. 
 
Now, there are (at least) two important choices to be made about how to implement the relativization of quantum states. First, we must choose between effective and inherent relationality: 

\begin{itemize}

\item \textbf{Effective quantum relationality:} In scenarios such as the Wigner's Friend case there may be emergent relational quantum descriptions, but there also exists an underlying `absolute'  quantum state which is not relativized to anything. 

\item \textbf{Inherent quantum relationality:} Quantum mechanics is inherently relational: the formalism is designed to describe relations between systems, so we cannot posit `absolute'  quantum states which are not relativized to anything.

\end{itemize}

This choice is about the meaning of the quantum formalism.   Effective approaches (e.g. Everett and de Broglie-Bohm) maintain that the quantum formalism operates by default in an `absolute' mode where  quantum states are assigned to systems without relativization, while also allowing that in certain regimes the theory may operate in a `relational' mode where we get effective relative quantum states relativized to something like a branch or a Bohmian particle.  By contrast, inherent approaches (e.g. RQM) take the Wigner's Friend paradox as evidence that quantum mechanics has \emph{only} a relational mode: what the quantum formalism describes is nothing other than relations between systems, so it is a category mistake to assign an absolute, non-relativized quantum state.

\vspace{2mm}

Second, we must choose between dynamical relationalism and fact-based relationalism:

\begin{itemize}

\item \textbf{Dynamical relationalism:} Quantum states are  relativized because the dynamics of a measurement interaction between two systems may depend on properties of both systems. 

\item  \textbf{Fact-based relationalism:} Quantum states are  relativized because facts are  relative.  

\end{itemize}

This choice is about the object of relativization. Fact-based approaches (e.g. Orthodox RQM) maintain that it is facts themselves that are relativized to observers;  so they defuse the tension exhibited in the Wigner's Friend paradox by simply maintaining that the facts of reality relative to Alice are different from the facts of reality relative to Bob. Whereas dynamical approaches (e.g. Everett, de-Broglie Bohm) say that relational effects in quantum mechanics arise because  quantum systems partake in a complex global dynamics which sometimes entails that the result of a measurement depends on who is doing the measuring.  So they defuse the tension exhibited in the Wigner's Friend paradox by saying that there is no real contradiction present, since Alice's predictions based on the state $\psi_A$ are correct \emph{for measurements performed by Alice} and Bob's predictions based on the state $\psi_B$ are correct \emph{for measurements performed by Bob}. No facts are relativized; we are simply making different predictions for two different possible interactions.

 Here it is helpful to distinguish between two meanings of the word `state':  as \cite{hughes1989structure} puts it,  `in classical mechanics the state could be said to have both a descriptive and a dispositional aspect.'  That is, states are sometimes used in a  `dispositional' sense, to characterize the possible future interactions the system may engage in  - for example, predicting the outcomes that will be observed if the system is measured. I will refer to this kind of state as an `interaction state.'  But states are also sometimes used in a `descriptive' sense, to characterize the actual current categorical properties of a system. I will refer to this kind of state as an `occurrent condition.' That is, interaction states describe modal facts (about possible future events) while occurrent conditions describe non-modal facts (about what is currently true).  

 With this terminology in hand, we can more clearly state the nature of the disagreement between dynamical and fact-based relational approaches. Both types of approach agree that due to the Wigner's Friend scenario, we must accept that in a quantum context, \emph{interaction states} are sometimes relativized to observers.  Fact-based relational approaches then stipulate that occurrent conditions are identical with or can be read directly off interaction states, so they infer from the relativization of interaction states that occurrent conditions or `facts' must also be relativized.  
 
 By contrast, dynamically relational approaches do not read occurrent conditions   directly off interaction states, so they do not infer from the relativization of interaction states that occurrent conditions or `facts' must be relativized\footnote{Note that the way in which we  define the `interaction state' will necessarily depend on the way in which measurements and interactions are identified and individuated, so  the question of whether interaction states are relativized or not will depend on our chosen way of individuating measurements. This makes sense in a dynamically relational view: if the relativization of interaction states is understood as just a convenient way of extracting local content out of a complex global dynamics, then it is to be expected that we will obtain different descriptions of the local dynamics when we use different descriptive conventions. For simplicity, in this article I will choose the convention that the identity conditions for measurements match the standard mathematical formalism of quantum mechanics, so Alice and Bob are considered to be making `the same measurement' if their respective operations performed at or shortly before the time of the measurement are represented in the formalism using the same set of mathematical operators.}. After all, the interaction state of a system is, by construction, about its relation to external systems with which it might interact, so it is makes sense to expect that interaction states may sometimes be relativized to such external systems. Whereas occurrent conditions do not describe relations to external systems, so this motivation does not apply to them and thus there is less reason to expect them to be relativized.

\vspace{2mm}

Evidently then we have various different options for relational approaches to quantum mechanics. The two choices above lead to a four-way classification of possible relational approaches: 

\begin{center}
\begin{tabular}{ c |  c  | c }
  & \textbf{Effective  } & \textbf{Inherent  } \\ \hline
\textbf{Fact-based}  & X  & Orthodox RQM  \\ 
& & Perspectival modal interpretations \\
& & Neo-Copenhagen/perspectival approaches
\\ \hline
 \textbf{Dynamical}  & Everett interpretation  & RQM+CPL \\
&   de Broglie-Bohm &   
\end{tabular}
\end{center}

\vspace{2mm}

 Let us consider these options in turn. 

\paragraph{Effective-fact} It does not appear that any existing approach falls into this category.  In principle it would be possible to design such 
 a view; this would require positing that some but not all physical facts are relative, so we could say there exist some non-relativized quantum states while also maintaining that in the Wigner's Friend scenario Alice and Bob have differing relative facts about the system $S$. 
 
 However, this kind of view looks quite unstable; for once we admit the existence of an underlying bedrock of absolute facts, it's tempting to say that only the `absolute facts' are really facts, and the supposed `relative facts' are really just absolute facts about relations. For example, rather than saying that `it is a fact relative to Alice that the outcome of her measurement was 0' we might just say that there is an absolute fact about the relation between Alice and the measured system.  This instability is  probably the reason why fact-based relational approaches often end up insisting that  \emph{all} meaningful physical facts must be relativized. Thus although it might be possible to develop an effective-fact approach, it's not clear there is any strong motivation for doing so.

\paragraph{Inherent-fact}
The  inherent-fact category is quite popular, subsuming many existing relational and perspectival approaches to quantum mechanics. 

However,  it should be noted that fact-based relationalism is going much further than necessary to resolve the Wigner's Friend paradox. 
For the impetus for relativization in that scenario is simply  the fact that if we want first-person universality to be upheld, Alice's assignation of the state $\psi_A$ must give the right predictions for Alice's measurements on $S$, and Bob's assignation of the state $\psi_B$ must give the right predictions for Bob's measurements on $S$. We can certainly still say that there is an absolute fact about the outcome of Alice's measurement on $S$, while also noting that the  dynamics of unitary quantum mecahnics work in such a way that the specific outcome obtained by Alice is not always dynamically relevant to Bob's interactions with  $S$, thus explaining why $\psi_B$ is the correct state to describe his interactions with $S$. Therefore what the Wigner's Friend scenario primarily demonstrates is  that  \emph{dynamics}, but not necessarily facts, must be relativized to observers

Moreover, when we look at the differences between classical and quantum physics, we can see that this kind of dynamical relativization arises in a very natural way in the quantum context. First, recall that in classical physics we typically expect that the facts about how two systems will interact will be fully determined by the sum of  their individual intrinsic  properties,  so the modal, dynamical facts about  a classical system follow directly from its non-modal properties, and thus the interaction state and occurrent condition of a system are in effect one and the same. And indeed, in the classical context  the word `state' is used to refer to both interaction states and occurrent conditions without further discrimination. 
 
 But in a theory like quantum mechanics where entanglement phenomena play a major role, the facts about how two systems will interact are not always fully determined by the sum of their individual intrinsic non-modal properties. Thus when we try to represent such a dynamics using individualized interaction states, these interaction states may no longer be determined uniquely by the occurrent condition of the system: we sometimes have to take into account the way in which the system is entangled with the systems it is interacting with. For example, this is exactly what happens in an Everettian or Bohmian description of the Wigner's friend scenario, where we end up with different effective interaction states of $S$ relative to Alice and Bob respectively, because of the different ways they are entangled with $S$. Thus the kind of relativization seen in the Wigner's Friend scenario is an inevitable consequence of the global, holistic nature of the quantum dynamics, so there is no need to postulate the relativization of \emph{facts} to explain it.

It should also be emphasized that many arguments for inherent relationalism work just as well for the dynamical approach as  fact-based approaches. For example, one common line of argument notes that quantum mechanics has been developed by physically embodied observers in order to describe the outcomes of their measurements, sand we should not expect this methodology to produce an absolute, observer-independent description of reality: quantum mechanics is necessarily a description of how reality impinges on embodied observers \citep{Fuchs_2014,Healey2012-HEAQTA}. But this is equally true if we think of quantum states as primarily descriptions of dynamics rather than `facts' or occurrent conditions: quantum mechanics has been designed to describe how embodied observers interact with other physical systems, so quantum states are inherently relational because they are designed to encode the dynamics of the different possible interactions that a  physical system can have with other systems. There is no need to appeal to relativized \emph{facts}   to make sense of the inherent relationality resulting from our epistemic situation, and thus again, fact-based relational approaches go further than the motivating idea really necessitates.

Of course, one might still be drawn to fact-based relationalism due to other  philosophical commitments. But it should be emphasized that this is a step which embroils us in significant epistemic and metaphysical complexities, so it should not be taken lightly - I will discuss some of these issues further in section \ref{compare}.  Moreover, as argued in  \cite{adlam2024moderatephysicalperspectivalism}, many of the philosophical intuitions which may appear to point towards a radical form of fact-based relationalism are actually more compatible with a moderate approach which maintains the existence of absolute facts, so it is not clear that general philosophical considerations can justify choosing fact-based relationalism over dynamical relationalism.

\paragraph{Effective-dynamical} This category is also quite popular, including well-established approaches like the Everett interpretation and the de Broglie-Bohm approach.   However, when we consider the way in which relationalism is instantiated in these views, there is an interesting conceptual tension. 

The issue is that these effective-dynamical approaches are committed to the view that quantum mechanics has both an `absolute' mode and a `relational' mode. That is, they suggest that the quantum formalism  by default describes systems in an absolute, non-relativized manner; but they also allow that we may obtain effective relational descriptions in certain contexts, e.g. the relative states of Everett or the conditional wavefunctions of de Broglie-Bohm. Moreover, these two modes are conceptually very different. In the absolute mode, quantum states are often interpreted as descriptions of occurrent conditions - for example,  some versions of the Everett approach take it that the  ontological content of the theory is given by the non-relativized universal quantum state \citep{carroll2018maddogeverettianismquantummechanics,Vaidman2019-VAIOOT}. But in the relational mode, a system may be assigned incompatible states relative to different observers, so these relative states cannot both be complete and literal descriptions of the occurrent condition of the system\footnote{Obviously in a fact-based relational approach, two incompatible states can both be complete and literal descriptions of the occurrent condition of a system, relative to different observers; but in a dynamically relational approach we do not relativize facts, so in this setting two incompatible relative states cannot both be complete and literal descriptions of occurrent conditions.}: often it seems more natural to understand them as interaction states, encoding modal, dispositional facts about what would occur if the system were to interact with the relevant observers. Yet we use exactly the same formalism for both the absolute and relational modes; one and the same mathematical object, the `quantum state,' describes both occurrent conditions, in the absolute mode, and interaction states, in the relational mode.  

This should worry us: positing two modes of the theory which are conceptually very different but formally identical might be taken as a sign of some kind of conceptual confusion. Moreover,  since any observation we can make necessarily involves an interaction between an observer and another physical system, it appears that all our direct experience of quantum mechanics is of its relational mode. For example, in the Everett interpretation,  no observer internal to the universe will ever access the absolute global quantum state; rather each observer only sees  effective relative states relativized to their current branch. So it is perhaps unclear   why we should believe that the formalism \emph{has} an absolute mode, in addition to the relational mode which pertains to our  actual observations. We cannot directly interact with the theory in its absolute mode, and moreover extending the theory to the absolute mode is not merely a simple extrapolation, but rather an entirely new application of the theory to a regime in which quantum states have a very different meaning. 

There are also other reasons to be suspicious of  the `absolute' states appearing in effective relational views; for approaches which posit absolute quantum states typically also posit an absolute global quantum state of the universe as a whole, and this often leads to tension with relativistic ideas. For example, this is one of the main difficulties for the de Broglie-Bohm approach - the  Bohmian absolute global state is defined on and evolves relative to a preferred foliation of spacetime, which requires us to posit a preferred reference frame, thus making  Bohmian mechanics look inconsistent with relativity. There exist various proposed solutions to this problem \citep{Galvan_2015,D_rr_2014}, but at this time none commands universal agreement. Whereas if we think from a more relational standpoint, it seems unnecessary to postulate this global quantum state at all: all the relevant empirical information is contained within a patchwork of local relativized quantum states, so we don't need to take the global quantum state too literally.   

On these grounds, it is tempting to speculate that quantum mechanics has \emph{only} the relational mode:  quantum states always encode relational interaction states, so they are just not the kind of thing which could possibly describe  occurrent conditions. As \cite{hughes1989structure} puts it, there are compelling reasons to think that  `in quantum theory the descriptive aspect disappears and we are left with the
dispositional aspect alone.' From this point of view, it appears that  the effective-dynamical approaches have simply misunderstood the nature of the quantum formalism: they have taken a theory developed specifically as a characterization of the way in which  creatures like us can interact with the world, and then erroneously interpreted it as an objective, observer-independent characterisation of reality. Therefore the effective-dynamical approach is in some ways quite unstable; once we acknowledge the existence of some kind of dynamical relationalism, there is a pressure towards acknowledging that the quantum state is by its very nature relational, and the inherent-dynamical approach is the logical endpoint of this progression.

\paragraph{Inherent-Dynamical}

 We have seen that although there are compelling reasons to think that the correct interpretation of the quantum formalism must involve some kind of relationalism,  neither scientific nor philosophical arguments provide a clear motivation for going to the extreme of fact-based relationalism; dynamical relationalism can serve equally well, while encountering fewer epistemic difficulties.  But we have also seen that most existing dynamical approaches fall within the effective-dynamical category, whereas there are some indications that the inherent-dynamical category may be more natural. And yet there are not many extant inherent-dynamical approaches; this suggests that further exploration of the inherent-dynamical category could be a promising way forward for the interpretation of quantum mechanics.

 \subsection{Modal Interpretations}

     Another way of motivating  inherent-dynamical approaches is based on considering how such a view would fit into the spectrum of modal interpretations. To see the connection, note that the distinction between interaction states and occurrent conditions  is similar to a distinction appearing in the context of modal interpretations. There, the quantum state is interpreted as a  `dynamical state,' which means it encodes information about `which physical properties the system may possess, and what probabilities are attached to these possibilities' \citep{sep-qm-modal}, and  separately, systems also have a `value state' which represents `what actually is the case, that is, all the system’s physical properties that are sharply defined at the instant in question' \citep{sep-qm-modal}. Thus dynamical states, like interaction states, encode modal facts, while value states, like occurrent conditions, encode non-modal facts. So inherent-dynamical views clearly have something in common with modal interpretations: both approaches identify the quantum state as modal, while also postulating some additional non-modal facts. 

However, the existing modal interpretations are not inherent-dynamical views as I have defined that term.  First, existing modal interpretations usually understand the dynamical state of a system $S$ at time $t$ as pertaining to what may be the case about $S$ \emph{at the time $t$}, i.e. the role of the quantum state of $S$ at a given time is to prescribe probabilities for the value state of $S$ at that very time \citep{dieks2007probabilitymodalinterpretationsquantum}. Whereas I have suggested that the modal nature of the quantum   state of $S$ at time $t$ pertains primarily to what may occur during \emph{future interactions} involving $S$, and therefore I am using the term   `interaction state' rather than `dynamical state'  to emphasize the future-oriented, relational nature of the quantum state. This difference is important, because  many modal interpretations  assume that dynamical states and hence quantum states can be assigned in an absolute way -  for example, some such approaches allow us to assign a quantum  state to the whole universe \citep{Bacciagaluppi1999-BACDFM}, whereas clearly it would not make sense to assign an \emph{interaction} state to the whole universe. So evidently  modal approaches of this kind will not fall into the category of inherent quantum relationalism.

Second, the existing modal interpretations  can be separated two main classes:  the `absolute' modal interpretations, which postulate that the  quantum state and value state of a given system are both absolute, observer-independent facts   \citep{DIEKS1989439,Vermaas1995-VERTMI,Bacciagaluppi1999-BACDFM}, and   the `perspectival' modal interpretations, which postulate that both the  quantum state and  the value state of a given system are  relativized to another physical system \citep{Bene2001-BENAPV,kochen1985new}.  The absolute modal interpretations are not relational at all, and meanwhile the perspectival modal interpretations implement fact-based relationalism, since the value states themselves  are relativized; so neither type of modal interpretation will typically fall into the category of inherent-dynamical relationalism. 

By contrast, an inherent-dynamical approach splits the difference, positing that the  quantum  state   is relativized, but the  occurrent condition  is not relativized. The closest analogue of this within the modal framework would be a view which  relativizes only dynamical states and not value states, but such an approach has not been  significantly explored in the literature on modal interpretations. There is an obvious reason for this:  traditionally an important goal of modal interpretations has been to provide a prescription which allows us to calculate  probabilities for the value state  directly from a quantum state \citep{dieks2007probabilitymodalinterpretationsquantum,DIEKS1989439}, whereas such a calculation will not in general be possible if we allow that there may be multiple incompatible quantum states associated with a given value state. But as noted   in section \ref{intro}, there are good conceptual motivations for the inherent-dynamical approach, so it is perhaps worth taking  this kind of intermediate modal approach seriously. Thus in addition to thinking of inherent-dynamical views as  filling a gap in the spectrum of existing relational approaches, we might also think of them as filling a gap in the spectrum of existing modal approaches.

\section{Orthodox RQM and RQM+CPL \label{orthodoxCPL}}

We have seen that there are some good reasons to think that inherent-dynamical approaches could be a promising way forward for the interpretation of quantum mechanics. However, there are not many extant examples of inherent-dynamical approaches. In this section, I will argue that RQM+CPL  is naturally understood as an inherent-dynamical view, and thus insofar as the inherent-dynamical approach is both promising and underexplored, RQM+CPL is interesting in its own right as a plausible interpretation of quantum mechanics. 

Let us begin with Relational Quantum Mechanics (RQM). This view, originated by Carlo Rovelli \citep{1996cr}, is a form of inherent quantum relationalism - other examples include perspectival modal interpretations \citep{Bene2001-BENAPV,kochen1985new} perspectival approaches \citep{https://doi.org/10.48550/arxiv.1801.09307} and neo-Copenhagen approaches \citep{brukner2015quantum}. Such approaches bring the relational aspects of quantum mechanics to the foreground, maintaining that the quantum formalism is by its very nature a description of relations between systems, so it cannot be applied to describe an individual system alone. 

RQM is distinguished within this class by its emphasis on avoiding anthropocentricism. Many relational or perspectival approaches posit that quantum descriptions are  relativized to `observers,' referring specifically to conscious, and/or human, and/or macroscopic systems.  Whereas RQM posits that quantum descriptions can be relativized to any physical system, so conscious, human, or macroscopic observers do not play any special role in the formulation of the theory. Thus  in this article, I will take RQM to be defined by the following postulates:

\begin{enumerate}

\item \textbf{No Anthropocentricism:} All physical systems can play the role of `observers' to which quantum descriptions may be relativized. 

\item \textbf{Inherent relationalism:} All quantum descriptions are relativized to physical systems, and the correct quantum description of a system $S_0$ relative to a system $S_1$ is not always the same as the correct quantum description of $S_0$ relative to a different system $S_2$.

\item \textbf{Measurements:} In any physical interaction between two systems $S_0, S_1$ a variable of  $S_0$ becomes definite relative to $S_1$; this can be described as a `measurement' performed by $S_1$ on $S_0$, and the value that becomes definite is the `outcome' of the measurement. 

\item \textbf{First-person universality of quantum mechanics:} For any observer (i.e. physical system), the correct predictions for the outcomes of a `measurement' performed by that observer are always obtained by applying a quantum description to the rest of the world at a suitable point in the past and evolving unitarily up until the time of the measurement interaction, then applying the Born rule. 

\item \textbf{Single-world:} Observers (i.e. physical systems) do not undergo branching upon measurement: when two systems $S_1, S_2$ undergo a measurement interaction and a variable $V$ of  $S_1$ becomes definite relative to   $S_2$, subsequently there  exists only one  copy of the system $S_2$, relative to which there exists only one value of the variable $V$.

\end{enumerate}

Now,  these postulates offer only a sketch of what the final view will look like. In particular, as pointed out by \cite{Mucino2022-MUCARQ} and \cite{https://doi.org/10.48550/arxiv.2107.03513}, we still require a more detailed account of the relationship between a general quantum interaction and the corresponding `measurement.' But let us suppose for now that these technical issues can be resolved. Even so, there remains an important conceptual question:  since RQM tells us that variables become definite `relative to a system' during a quantum measurement, what can we say about the relationships between variables relativized to different systems? Do there exist any absolute, observer-independent connections between variables relative to one system and variables relative to another system, or does each system have its own reality which is disconnected from other systems? 

In this article, I will compare two different ways of answering this question.  First of all, the response given within Orthodox RQM. Here it is important to note that  Orthodox RQM is usually presented as a form of fact-based relationalism, so in this context the relativization of quantum states is understood to reflect  the relativization of physical facts themselves. Indeed, it is commonly suggested that in Orthodox RQM   \emph{all} physical facts must be relativized to physical systems, and thus it is meaningless to make any statement about physical reality without relativizing that statement to some physical system \citep{pittphilsci19664}, which means we cannot even pose questions about the relations between facts relative to one system and facts relative to any other system. We are sometimes permitted to say that there exists a relationship between Alice's observations and Bob's observations \emph{relative to the perspective of some third system}, but it simply does not make sense in this view to ask  whether Alice's observed outcomes as she herself experiences them  bear any relation to Bob's observed outcomes as he himself experiences them. Indeed, this is the way in which Orthodox RQM proposes to resolve the Wigner's Friend paradox: Alice and Bob can assign different quantum states to the same system because they each have their own set of relative facts, and we need not  worry about conflicts between their relative facts because Orthodox RQM denies the meaningfulness of comparisons between relative facts associated with different systems. 
 
Alternatively, we have the response given by \cite{https://doi.org/10.48550/arxiv.2203.13342}, who suggest that RQM should explicitly address the question of the relationship between variables relativized to different systems by adding an extra postulate to the five postulates above: 

\begin{enumerate}[resume]
 
 \item \textbf{Cross-Perspective Links (CPL):} In a scenario where some observer Alice measures a variable V of a system S, then provided that Alice does not undergo any interactions which destroy the information about V stored in Alice’s physical variables, if Bob subsequently measures the physical variable representing Alice’s information about the variable V, then Bob’s measurement result will match Alice’s measurement result.\footnote{This postulate follows the standard convention in RQM of using the word `observer' to refer generally to any physical system to which a quantum state can be relativized; it need not be assumed that Alice and/or Bob are conscious, human or macroscopic.}

\end{enumerate}

In this article, I will understand this postulate to be asserting the  existence of a relationship between Alice's perspective and Bob's perspective in an absolute sense. Thus interpreted, the CPL postulate commits us to the existence of some facts about physical reality which  are not relativized.

\cite{rovelli2024alicesciencefriendsrelational} notes that one could formulate a weaker version of this postulate as follows: 
 
\begin{itemize} 
 
 \item \textbf{Relativized Cross-Perspective Links (RCPL):}  For any physical system $S$,  in a scenario where it is the case that  \emph{relative to  S} some observer Alice measures a variable V of a system S, then provided that \emph{relative to S} Alice does not undergo any interactions which destroy the information about V stored in Alice’s physical variables, if it is the case that \emph{relative to S} some other observer Bob subsequently measures the physical variable representing Alice’s information about the variable V, then Bob’s measurement result  \emph{relative to  S} will match Alice’s measurement result  \emph{relative to S}.

\end{itemize}

The RCPL postulate does not commit us to the existence of facts about physical reality which are not relativized, and thus it is plausible that RCPL is upheld within Orthodox RQM. However, I will argue in section \ref{compare} that RCPL fails to solve the problems that CPL was initially intended to address, so there are good reasons to prefer the absolute CPL postulate. For now, I will focus on a version of RQM which uses the absolute version of the postulate, and  I will refer to this version of RQM  as `RQM+CPL.' 

\subsection{RQM+CPL }

 Given the characterisation of RQM in the previous section, it is evident from the second postulate that both RQM and RQM+CPL  both fall into the category of inherent quantum relationalism. And as we have already noted, Orthodox RQM is usually presented as a form of fact-based relationalism. On the other hand, RQM+CPL is naturally understood as a form of dynamical relationalism. To see this, let us consider what RQM+CPL says about the Wigner's Friend scenario.

First of all, the first-person universality of quantum mechanics entails that after Alice's measurement, the superposition state $\psi_{SA}$ gives the correct predictions for any measurement performed on $S$ and/or $A$ by Bob. This means that $\psi_{SA}$ is the \emph{interaction} state of $S \otimes A$ relative to Bob, i.e. it correctly describes his interactions with $S \otimes A$, although it may not necessarily be a correct description for the interactions of other observers with $S \otimes A$.

Second, the CPL postulate ensures that even though $\psi_{SA}$ carries no record of the actual outcome of Alice's measurement, nonetheless   there exists a fact about the actual outcome of Alice's measurement, which Bob could in principle find out if he were to measure Alice in an appropriate basis, and any other observer could also find out if they were to measure Alice in an appropriate basis. So although $\psi_{SA}$ is the interaction state for Alice and $S$ relative to Bob, $\psi_{SA}$ does not encode all the facts about the occurrent condition of Alice and $S$. Thus the  interaction state of  $S \otimes A$ is relativized to Bob in this scenario, but the system also has an occurrent condition which is not relativized to Bob.

Now, one might worry that there is some inconsistency in this account; for surely  $\psi_{SA}$ cannot be the correct interaction state relative to Bob when it is possible in principle to make better predictions than those encoded in $\psi_{SA}$ for the outcome of Bob's measurement in one specific basis $B$, i.e. the basis which provides evidence about Alice's measurement outcome as specified by the CPL postulate. However,  if we repeat this experiment many times and have Bob perform tomographic measurements, the statistics he sees will always be as predicted by   $\psi_{SA}$, because the linearity of quantum mechanics entails that the expected relative frequencies for the Bob's measurement in basis $B$ are same as the expected relative frequencies for Alice's measurement on $S$. So rather than an inconsistency, here we have a scenario  where the definition of \emph{the} `(interaction) state' becomes blurry, even when relativized to a specific observer: there is an ambiguity about whether the `interaction state' should be defined in a statistical sense or with respect to the facts of a single case. This ambiguity is not a problem as long as we distinguish clearly between interaction states and occurrent conditions, as suggested by the dynamically relational approach. For then we can say that the interaction state is simply a convenient tool for extracting locally meaningful consequences for various possible interactions out of the complex global dynamics of quantum mechanics, and thus there is no fact of the matter about the `true'  interaction state:  we can happily acknowledge that both the statistical and single-case definitions may yield useful information. The important point  is that whatever choice we make, the CPL postulate does not lead to any empirical predictions which are in contradiction with standard quantum mechanics as applied by Bob.

    This description of the Wigner's Friend scenario suggests a natural way of understanding what RQM+CPL says about the world. First, it tells us that quantum states like $\psi_{SA}$ are always interaction states, i.e. they are tools for extracting information out of a complex global dynamics, and they are relativized because they describe a relation to an external system with which an interaction could occur. Second, it tells us that there exist underlying occurrent conditions distinct from these relativized interaction states, and observers can find out about one another's occurrent conditions under certain circumstances. Thus the relativization of quantum states is an inherent feature of  quantum mechanics, but this relativization does not pertain  primarily to facts;   like the dynamical relationalism appearing in Everett and de Broglie-Bohm, it is a natural result of the global structure of the quantum dynamics. Thus understood, RQM+CPL is clearly a form of inherent-dynamical relationalism. Of course, RQM+CPL is surely not the only possibility in this space, but it is useful as a working example of what such a thing would look like\footnote{It is possible that the de Broglie-Bohm approach could also be adapted in some ways to give an inherent-dynamical approach. For example,  if one adopts the version of the Bohmian approach in which the wavefunction is understood as modal or nomic, and if one then adds the stipulation that there is no absolute global quantum state, rather just a patchwork of local relativized quantum states, then this would presumably be an inherent-dynamical approach. It would be interesting to develop this possibility further, and to compare it to RQM+CPL to see if the two approaches may converge.}. 

  So let us note some important features of RQM+CPL as they pertain to the nature of the relativization in quantum mechanics. First, we can see that the quantum state plays a specific kind of role in this approach, lying somewhat outside of the traditional `ontic vs epistemic' distinction \citep{HarriganSpekkens}. Here  quantum states are identified as interaction states, so they do not encode   occurrent conditions or `ontic states,'; but  they are also not merely subjective or epistemic. The quantum state  $\psi_{SA}$  describes concrete dynamical facts about what would happen if Bob measured $S$ and/or Alice, and it remains the correct interaction state regardless of whether Bob is capable of understanding it or writing it down. It is true that the interaction state relative to Bob can be affected by changes in Bob's knowledge - for example, if he learns Alice's measurement outcome, the interaction state of $S$ relative to him will change from $\psi_B$ to  $\psi_A$. But this is not because these states are \emph{descriptions} of Bob's knowledge; rather, the point is that gaining knowledge is a physical process which changes the way in which  Bob is entangled with $S$, and thus it changes the dynamics for his future interactions with $S$, so the interaction state of $S$ relative to him must change. Thus insofar as dynamics are a real, objective feature of the world, quantum  states are also a real, objective feature of the world in RQM+CPL, even though they are not `ontic states' in the usual sense.   

Second, we can see that the modal nature of the interaction state is important for the consistency of inherent-dynamical views.  For although RQM+CPL allows that a  system $S$ may have incompatible interaction  states at a time $t$ relative to Alice and Bob respectively, at most one of Alice and Bob can \emph{actually} measure $S$ at that time $t$, so we will only ever need to invoke one  interaction  state to determine the outcome of the measurement that actually takes place. This is why it is possible within RQM+CPL to relativize  interaction  states and yet maintain the existence of absolute occurrent conditions: it doesn't matter if different  interaction  states disagree with each other, since interaction states only describe possibilities, and at most one  such state is ever relevant for determining what actually occurs at a given time. This high-level argument  gives good reasons to think that inherent-dynamical views like RQM+CPL will not encounter any insurmountable inconsistencies.

\subsection{Ontology}

Since RQM+CPL posits some absolute, non-relativized facts, it is natural to ask what these facts might look like. The CPL postulate mandates that there exist absolute, non-relativized facts about the observed outcomes of measurements and the relationships between such outcomes, but otherwise there is considerable freedom to imagine different possibilities. In this article I will  not assume any particular ontology for RQM+CPL, but will use this term to refer generally to any approach which obeys the five postulates of RQM plus the CPL postulate, with the understanding that this might be compatible with a variety of ways of thinking about the underlying `absolute' facts. 

Here we should take note of some important consequences of the commitments involved in an inherent-dynamical approach. In such a picture, the quantum formalism is understood as a way of organizing our knowledge about a small subset of the underlying absolute facts from a particular embodied standpoint, and predicting facts that lie in the future of that standpoint. The quantum formalism cannot be used to directly characterize the distribution of absolute facts as a whole, because that would require us to ascribe quantum states in a non-relativized way, which does not make sense if we believe that quantum mechanics is inherently relational. Thus although inherent-dynamical approaches do postulate some absolute, non-relativized facts, they also tell us that we should not try to use the standard formalism of quantum mechanics to express or derive these facts. In this context, the appropriate approach to questions of ontology is not to try to derive a set of absolute facts using the quantum formalism directly, but rather to seek  a different kind of description of the underlying facts and then show mathematically  how quantum mechanics emerges as an appropriate relational description in a relevant limit.
 
So for example, although there are  certainly some unanswered questions about when and how the values of variables become definite in RQM, in the context of RQM+CPL it is clear that we should not   to try to derive answers to these questions directly from the formalism of standard quantum mechanics. This can  be seen as follows. In the context of inherent quantum relationalism, quantum states always   describe relations to external systems, so the  quantum description of the world relative to a system $S$ cannot  include a quantum state for $S$. Thus this description cannot tell us anything about when and how $S$ will interact with some other system $S'$, since $S$ itself does not feature in the description. The same applies for a description relative to $S'$. Of course we can always move to a quantum description of $S$ and $S'$ relative to some third system $S''$ instead; but  in RQM+CPL the the quantum description of $S$ and $S'$ relative to $S''$ is merely an interaction state describing the dynamics that would take effect if $S''$ were to interact with $S$ and/or $S'$, so it does not encode the occurrent condition of $S$ and $S'$. Thus although this description might superficially look as if it is describing an interaction between $S$ and $S'$, it cannot directly inform us about when and how values become definite for $S$ and $S'$, since that is a feature of their occurrent condition. So there is no quantum description which can directly tell us how and when  values become definite relative to $S$ and $S'$; the answers to such questions would inevitably have to fall outside of the standard quantum formalism, and the goal would be to answer these questions in such a way that we recover quantum mechanics as an effective relational description of the process of values becoming definite in some appropriate regimes. 

Evidently then, writing down a concrete ontology for an inherent-dynamical approach  would not be straightforward, and would likely require appealing to physics beyond standard non-relativistic quantum mechanics. On the other hand, the postulation of underlying absolute facts does at least allow us to clearly formulate this mandate and consider responses to it, whereas in the context of inherent-fact approaches it is difficult to even formulate the problem. For example,  critical discussions of Orthodox RQM reveal a clear need to  make precise the claim that variables become definite in interactions \citep{Mucino2022-MUCARQ,https://doi.org/10.48550/arxiv.2107.03513},  but since Orthodox RQM forbids us from postulating any underlying absolute facts, it is quite difficult to properly state this problem and say how it can be addressed. Thus although ontology is certainly an outstanding issue in the context of RQM+CPL, it is a positive sign that we can state this issue as a concrete research problem which can be addressed with mathematical tools and which has clear criteria for success.

\section{Orthodox RQM vs RQM+CPL \label{compare}}

I have argued that RQM+CPL provides a useful example of an inherent-dynamical relational approach to quantum mechanics, thus giving a clear motivation to explore the view further. In this section, I will make a direct comparison between RQM+CPL and Orthodox RQM, arguing that RQM+CPL  has some important advantages over the Orthodox version.  Note that although  I will focus on Orthodox RQM here for definiteness, many of these points are  also relevant to other inherent-fact approaches, so one might also understand these arguments as reasons to prefer inherent-dynamical approaches over inherent-fact approaches in general.

\subsection{Intersubjectivity \label{intersubjectivity}}

As described by \cite{https://doi.org/10.48550/arxiv.2203.13342}, the central motivation for the formulation of RQM+CPL  is  to ensure that there exist absolute, observer-independent connections between the relative facts perceived by different observers.   \cite{https://doi.org/10.48550/arxiv.2203.16278} argues that this is important because  the practice of science relies crucially on the sharing of information about measurement outcomes between different observers, and therefore any scientific theory which denies that measurement outcomes can be shared between different observers is empirically incoherent.

In particular, we saw in section \ref{motivation} that the  first-person universality of   quantum mechanics is an important founding premise of RQM, since one main motivation for relativizing quantum descriptions is to allow us to say that even in a Wigner's Friend scenario, quantum   mechanics is a correct description of the observations of all observers. Thus in order to empirically confirm quantum mechanics as it is understood within RQM, we need to obtain empirical evidence in support of the hypothesis that quantum mechanics describes the perspectives of all observers. This requires obtaining evidence about the perspectives of several different observers, verifying that they all exhibit regularities consistent with quantum mechanics, and then performing induction to arrive at the conclusion that quantum mechanics applies to the perspective of all observers. 

But Orthodox RQM tells us that it is not possible for any one observer to obtain information about the perspectives of several different observers - each observer has access only to their own perspective, and  indeed it is not even meaningful to make comparisons between two different perspectives. Yet data from one perspective  clearly cannot provide us with  adequate evidence to justify believing the hypothesis that quantum mechanics applies to \emph{all} perspectives - we cannot perform induction based on a single data point!  So if Orthodox RQM is correct,  we cannot empirically confirm quantum mechanics \emph{as it is understood within RQM}; and of course if we cannot empirically confirm quantum mechanics then we  have no good reason to believe in Orthodox RQM  itself,  so the view  is self-undermining. 

By contrast, RQM+CPL explicitly assures us that it is possible for an observer to obtain reliable information about the perspectives of other observers via physical interactions with them. So although it of course remains true in RQM+CPL that each observer exists within  only one perspective at any one time, nonetheless RQM+CPL affirms that it is possible for such an observer to obtain data about the content of many different perspectives, and thus it affirms that observers are able to obtain appropriate empirical evidence which can allow them to perform an inductive inference to the conclusion that quantum mechanics correctly describes all perspectives.  Thus RQM+CPL offers a coherent account of the epistemology of quantum mechanics as it is understood within RQM, so it is does not have the same self-undermining problems as Orthodox RQM.
 
 In response to this worry, \cite{rovelli2024alicesciencefriendsrelational} points out that although the absolute version of CPL cannot be true within Orthodox RQM nonetheless the weaker `Relativized Cross-Perspective Links (RCPL)' holds within Orthodox RQM:  if Bob asks Alice about her outcome and hears her saying that she obtained some outcome $X$, then all his subsequent interactions will be consistent with this value $X$, so from his point of view it looks as if he knows the `true' value of the outcome. But it must be emphasized that all of this is a fact about Bob's perspective only: there is no possible way for Bob to learn anything about what Alice's perspective looks like, from \emph{her} point of view.  So the weaker postulate RCPL does not resolve the epistemic problem I have described, because if Bob cannot learn anything about Alice's perspective he can have no reason to think that it obeys regularities similar to those witnessed within his own perspective, and therefore he cannot obtain any evidence to support the hypothesis that quantum mechanics applies within perspectives other than his own. 
 
 Now  as emphasized by \cite{articleriedel}, pursuing Orthodox RQM to its logical conclusion seems to suggest that the facts about which relative facts hold  relative to one observer must themselves be relativized to some other observer. Thus one might try to overcome these epistemic objections by pointing out that although Orthodox RQM tells us that Bob cannot possibly learn about facts relative to Alice simpliciter, it does allow that Bob can learn about facts relative to Alice \emph{relative to him}. However, this  does  not help the epistemic situation, because it remains the case that in this picture a given observer, Bob can only ever find out about facts such that the chain of relativization ultimately terminates in \emph{his} perspective - e.g.  `The value of V is v, relative to Bob,' and `The value of V is v, relative to Alice, relative to Bob,' and `The value of V is v, relative to Alice, relative to Chidi, relative to Bob.' So Bob can perhaps verify that the `the set of all relational descriptions terminating in Bob' conform to the predictions of quantum mechanics\footnote{In fact, it is not entirely clear that Bob can even verify this much in Orthodox RQM - see the worries about knowledge of one's own observations discussed in section \ref{oneself} and also the concerns about the persistence of observers over time as discussed in \cite{https://doi.org/10.48550/arxiv.2203.16278}.}, but he cannot have any information about  `the set of all relational descriptions terminating in X,' for any X other than Bob. 
 
 And yet unless we are willing to say that  Orthodox RQM is really just solipsism, it seems that if Bob believes Orthodox RQM then he is committed to the existence of  relational descriptions where the sequence of relativizations does \emph{not} terminate in his perspective, and to the claim that these relational descriptions also conform to the predictions of RQM. That is, Bob is supposed to believe `For \emph{all} X, the set of all relational descriptions terminating in X conforms to the predictions of quantum mechanics.' And yet he can only test this claim for one value of X, so he cannot possibly have enough evidence to justify an inductive inference to such a conclusion. By construction, the theory denies Bob exactly the kind of evidence he would need to empirically confirm it, i.e. evidence about anything at all that is not ultimately relativized to him. Thus in order to resolve the epistemic difficulties here,  we need the absolute CPL postulate, not the weaker RCPL postulate. 
 
Now, proponents of Orthodox RQM may object to these epistemic arguments on the grounds that it does not even make sense within Orthodox RQM to ask whether Bob knows the true value of Alice's outcome or Alice's relative facts, since Orthodox RQM does not permit us to make comparisons between perspectives. So for example, the statement `Alice's actual outcome matches Bob's actual outcome, in an absolute sense' simply cannot be assigned a truth value within Orthodox RQM - it is not true, but it is not false either. However, this is not a counterargument to the points made above. The upshot of these epistemic considerations is that  in order for Bob to rationally  use his own observations to perform an inference to the conclusion  that quantum mechanics applies within all perspectives,  he needs to believe that he has some means of using his observations to gain  information about other perspectives, or about relative descriptions that do not terminate in him; that is, he must be able to say that statements of the form `Alice's actual outcome matches Bob's actual outcome, in an absolute sense,' are sometimes \emph{true}, or probably true. It is not enough to say such statements  are `not false,' because the assertion that  statements of this kind simply do not have a truth value is no use whatsoever to Bob if he is trying to understand what his observations  can tell him about regularities occurring beyond his own perspective. So we cannot resolve these epistemic problems just by reiterating the meaninglessness of comparing perspectives within Orthodox RQM: the fact that such comparisons are meaningless is precisely the problem at hand!

\subsection{Ownership of experiences}

Another significant reason  for preferring RQM+CPL  over Orthodox RQM is the common-sense idea that the final authority on an experience is the person to whom that experience belongs. That is, common sense suggests that if Alice performs a measurement and experiences it as having outcome $O_1$, while Bob believes that the measurement had some other outcome $O_2$ or that it had no definite outcome at all, then we surely need not say that Alice and Bob have an equal claim to the truth: Alice cannot be mistaken about the content of her own experience (or at least, she cannot be mistaken at the time of that experience, although she could perhaps remember wrongly afterwards). Thus common sense suggests that Bob must be in the wrong here: he has made an incorrect inference about someone else's experience, perhaps because he is incorrectly conflating interaction states with occurrent conditions. 

And this idea is indeed upheld in RQM+CPL, where Alice's occurrent condition includes facts about what she has actually observed, which remain true regardless of what her interaction state relative to Bob is like. On the other hand, Orthodox RQM denies this   idea: it is committed to the claim that a measurement performed by Alice can have one outcome relative to Alice and a different outcome relative to Bob, or it can have a definite outcome relative to Alice and yet fail to have a definite outcome relative to Bob. But on the face of it this is  an extremely strange claim. What could it possibly \emph{mean} to say that the outcome of the measurement performed by Alice is different relative to Bob than it is for Alice herself? If people have different views about the content of an experience, surely the person who actually had that experience is the ultimate authority on what the experience was?

Of course, we should not reject claims just because they are strange; but  we should reject them if no coherent meaning can be assigned to them. And in fact, attempts to elucidate the meaning of the claim that the facts about the outcome of Alice's measurement are different relative to Alice and to Bob have a tendency to collapse into dynamical relationalism, so it is not obvious that Orthodox RQM does in fact have a coherent meaning to assign to this claim. For example, in the Wigner's Friend scenario described in section \ref{intro}, the claim that `the outcome of Alice's measurement of S is indefinite relative to Bob'   is usually elucidated by just  reiterating  that  if Bob measures Alice and/or S, he will see statistics compatible with the superposition state $|\psi\rangle_{SA} = a|0 \rangle_S |0 \rangle_A + b|0 \rangle_S |0 \rangle_A$ rather than $| 0 \rangle_A | 0\rangle_S$ or  $|1 \rangle_A | 1\rangle_S$; but of course  this is also true in a dynamically relational version of RQM, so if this assertion about Bob's measurement outcomes is the whole content of the claim, then we have indeed collapsed into dynamical relationalism. 

Moreover, these questions about experience become even more challenging if we accept that, as argued by  \cite{articleriedel}, Orthodox RQM necessarily leads to an `iteration of relativization,' where the facts   about what relative facts are true relative to Alice must themselves always be relativized to another observer. For example, suppose that I perform a measurement and see an outcome $O$. One might think I can now say that it is true relative to me that the outcome of the measurement is $O$, but the iteration of relativization suggests that in fact, the proposition P, `It is true relative to me that the outcome is $O$,' can only be true relative to some other observer, and moreover the fact that  $P$ is true relative to that other observer can itself only be true relative to some \emph{other} observer, and so on ad infinitum. So what am I to think about the outcome that I actually remember observing - to whom is it relativized? What part of this infinite regress is reflected in my actual evidence?  Are there other copies of me relative to different observers? Which version of me am I?

This leads to a dilemma. If we say that there is necessarily one actual \emph{conscious} version of me who sees one specific outcome, i.e. the outcome that I remember, that  undermines the claim that there are no absolute facts about my outcome. But if we allow that  that there exist multiple conscious copies of me relative to various different observers, this will  add significantly to the epistemic difficulties involved in RQM, particularly if it is impossible even in principle to raise questions about which observer the experiences I currently remember are relativized to. Indeed,  in that case we will plausibly end up dealing with all of the same epistemic problems as the Everett interpretation \citep{AdlamEverett}, on top of the ones faced by RQM. And in fact, if we are willing to accept the existence of multiple conscious copies of an observer,  the Everett interpretation would surely  be a simpler and cleaner way to implement this!

So it appears that Orthodox RQM has quite significant problems with accounting for the connection between experience and physical reality in a coherent and natural way, whereas RQM+CPL avoids these problems by allowing that observers have occurrent conditions encoding absolute facts about what they have actually observed.

\subsection{Knowledge of oneself \label{oneself}}

Most presentations of RQM maintain that observers do not assign quantum states to themselves \citep{1996cr, pittphilsci19664, Robson2024-ROBRQM}. There is a good reason for this: if we allow that observers assign quantum  states to themselves, we are liable to run into inconsistencies. Indeed, critics of relational views have sometimes offered constructions which purport to demonstrate inconsistencies in such views,  but on examination these inconsistencies often arise from assuming erroneously that observers assign quantum states to themselves (for example, see  \cite{kastner2024conventionalquantumtheorydoes}). 
 
 But in the context of Orthodox RQM, the fact that observers do not assign quantum states to themselves has odd consequences. We cannot say that an observer has knowledge of her own quantum state relative to herself, since no such thing exists. Nor can she have knowledge of her own quantum state relative to other observers, since she has no access to facts relativized to other observers. Yet since Orthodox RQM does not posit occurrent conditions which are distinct from quantum states, she also cannot have knowledge of her own occurrent condition, even relative to herself. Therefore Orthodox RQM is apparently committed to saying that an observers cannot in fact   know \emph{anything} about her own condition, even relative to herself. 

Yet this seems possibly incoherent, or at least very hard to accept. I surely  know some things about my own condition, at least relative to me - indeed, in the spirit of \cite{descartes1993meditations}, one might say that my own condition is  the only thing I can know about with any certainty!  And this issue surely adds to Orthodox RQM's epistemic troubles - if I cannot know or have good reasons to believe that I have observed some measurement outcome, how can the process of doing science ever get started? 

One way for the proponents of Orthodox RQM to respond to this issue might involve the `iteration of relativity' posited by Riedel. Thus one might argue that although Alice cannot assign a state to herself, she can know relative facts of the following form `The state of Alice, relative to Bob, relative to Alice' - that is, Alice can have some knowledge of her own condition by means of interacting with other observers and using the results of those observations to make inferences about what her state is, relative to those other observers, relative to her. However, this would surely be a very odd way of understanding   knowledge of the self: it seems seems just empirically false to suggest that in order to know that she has observed a certain measurement outcome, Alice must interact with someone or something else and indirectly infer that relative to that other system she is in a state $| \psi_A \rangle$ which encodes the relevant measurement outcome. 

Moreover, in order for Alice to determine that it is true relative to her that relative to some other observer, Bob, her state is  $|\psi_A \rangle$, Alice would presumably have to know the results of some measurements on Bob in order to learn something about Bob's relative facts, relative to her. But then surely she can only know \emph{those} measurement results by interacting with some other system, e.g. Chidi, to infer that she is in a state $| \psi' \rangle$ relative to Chidi which encodes the appropriate results of her measurements on Bob, and so on. That is, this approach seems to trap Alice in an infinite regress where she can never actually use any measurement result to conclude anything about herself or the world more generally, since every measurement result must be backed up with an infinite series of interactions determining an infinite series of relative states. So this does not look like a viable way of understanding knowledge of the self.

By contrast, since  RQM+CPL distinguishes between interaction states and occurrent conditions, we do not encounter these problems. In RQM+CPL, of course  an observer does not assign a quantum  state to herself, since quantum states are interaction states and one cannot interact with oneself. But an observer can in principle have knowledge of her own occurrent condition, so she can have some kind of knowledge of the self even though she doesn't assign a quantum state to herself. So in a dynamically relational approach we are  able to avoid these kinds of consistency problems without denying that observers know something about their own condition, thus avoiding the apparent incoherence appearing in Orthodox RQM.

\section{Objections to RQM+CPL \label{objection}}

In this section, I discuss some possible objections to RQM+CPL, and to inherent-dynamical relationalism  more generally, and explain how they can be overcome.

\subsection{Regress of Relativization}

\cite{rovelli2024princetonseminarsphysicsphilosophy} offers the following argument which seems to be intended to rule out dynamical relationalism and push us towards a radical form of fact-based relationalism:

`Take the case that ... Friend interacts with S and measures S only if a quantum experiment gives him one outcome rather than another one. In Wigner’s account of the content of the room, there is a superposition between a state where S has a value for Friend, and a state where it doesn’t. Hence the fact itself that S has a value relative to Friend is only relative to something else: there is no non-relative fact of the matter about it.' Rovelli thus concludes that `Relativity iterates: that is, the above observation triggers an infinite regression that prevents us from making any meaningful absolute statement about what is the case.' 
 
However, this argument works  only if we assume that occurrent conditions can be read directly off interaction states, which is exactly what dynamical relationalism denies. To see this, let us apply Rovelli's argument to a specific example, where for consistency with previous examples I will rename Friend as Alice and Wigner as Bob. Suppose that Alice measures a system $Q$ originally prepared in the state $\frac{1}{\sqrt{2}} ( | 0 \rangle_Q + |1 \rangle_Q) $, and then if the outcome is $| 0 \rangle$, Alice measures system $S$ which is prepared in
the state $a | 0 \rangle + b |1 \rangle$ as in the original example in section \ref{intro}, so that from Bob's point of view 
Alice and $S$ are now in the entangled state $\psi_{SA} = a | 0 \rangle_S | 0 \rangle_A + b |1 \rangle_S | 1 \rangle_A$. Whereas if the outcome is   $|1 \rangle$ Alice does not measure $S$, so from Bob's point of view Alice and $S$ are in a product state $ (a | 0 \rangle_S + b |1 \rangle_S) \otimes |i \rangle_A$, where $|i \rangle_A$ is some fixed state of Alice.

 Rovelli's point here is that if Bob applies unitary quantum mechanics to describe this scenario, he will end up assigning a state of the following form to the system $Q \otimes S \otimes A$: 

 \[ \psi_0 = \frac{1}{\sqrt{2}} ( |0 \rangle_Q \otimes  \psi_{SA} + |1 \rangle_Q \otimes (a | 0 \rangle_S + b |1 \rangle_S)  \otimes  | i \rangle_A )   \]
 
Whereas if Bob believed that some variable of S definitely has a value relative to Alice, we might expect him to express this belief by writing down a state in which $A$ and $S$ are definitely in the entangled state $\psi_{SA}$, which might look something like the following:  

\[\psi_1 = \frac{1}{\sqrt{2}} ( | 0 \rangle_Q + |1 \rangle_Q) \otimes  \psi_{SA} \]

Now, if we want to maintain the first-person universality of unitary quantum mechanics, we are indeed obliged to say that when Bob measures the joint system $Q \otimes S \otimes A$, he will see results consistent with the state $\psi_0$ predicted by unitary quantum mechanics, rather than the alternative state $\psi_1$. That is, we must say that the \emph{interaction state} of $Q \otimes S \otimes A$ relative to Bob is $\psi_0$ and not $\psi_1$.

Rovelli is thus arguing that the fact that the (interaction)  state of $Q \otimes S \otimes A$  relative to Bob is $\psi_0$ entails that there is no fact of the matter about  whether or not $S$ has a value relative to Alice. However, this follows only if we assume that the occurrent condition of $Q \otimes S \otimes A$ is the same as or can be read directly off its interaction state relative to Bob.  In order to maintain the first-person universality of quantum mechanics in this scenario we need only  say that the state  $\psi_0$ correctly  encodes the outcomes that Bob will see for his measurements;  this does not in and of itself imply anything about whether $S$ has a value relative to Alice, or what that value is. That follows only if we make a further interpretative assumption to the effect that there cannot exist any facts about the system $Q \otimes S \otimes A$ which are not reflected in its quantum  state relative to Bob, i.e. occurrent conditions either do not exist or they are to be identified with or read directly off quantum  states. This is not true in any approach implementing dynamical relationalism rather than fact-based relationalism, so in such a context we cannot be forced into a regress of relativization.

\subsection{EWF Theorems \label{EWF}}

Recently various `Extended Wigner's Friend' theorems have reinvigorated the discussion of relationalism in quantum mechanics \citep{Bong_2020,ormrod2022nogo,2018qtcc}.   As with previous no-go theorems, the EWF theorems show that a certain set of assumptions jointly lead to predictions which are incompatible with the predictions of quantum mechanics. Thus an EWF theorem may be understood as  a demonstration that if we believe in the first-person universality of quantum mechanics, one of the assumptions in the corresponding set must be given up.

For example, in  the EWF theorem due to \cite{Bong_2020}, the three assumptions used to derive the contradictory predictions are `Locality,' `No-Superdeterminism,' and  `Absoluteness of Observed Events.' 
These assumptions are applied to a set of measurements performed by four observers, Alice, Bob, Chidi and Divya, where Chidi and Divya simply perform fixed measurements on systems $S_C, S_D$ respectively, and then Alice  can  choose to ask Chidi  the result of his measurement or to perform a non-commuting measurement of her own on the joint system made up of Chidi and $S_C$, and likewise mutatis mutandis for Bob and Divya. The discussion of this theorem has particularly  focused on the possibility of giving up Absoluteness of Observed Events, thus motivating the idea that `observed events' (i.e. measurement outcomes) are in some sense relativized in quantum mechanics. 

More specifically, the AOE assumption plays two distinct mathematical roles in the derivation of the contradictory predictions, by enforcing the following two requirements:

\begin{itemize}

\item \textbf{Single-world:} observers do not undergo branching upon measurement, so when Chidi performs a measurement then afterwards there is only one copy of Chidi and only one outcome of the measurement, $C$, relative to that copy of Chidi (and likewise mutatis mutandis for Alice, Bob and Divya). 

\item \textbf{Cross-Perspective Links:} when Alice asks Chidi about his measurement outcome and hears a value $A$, she learns the actual result of Chidi's measurement, i.e. we can assert an equality $A = C$ between the value Alice learns in this interaction and the value of Chidi's outcome relative to Chidi at the time of his measurement. 

\end{itemize}

 Now, the idea that the EWF theorems motivate giving up absoluteness of observed events  might look like a threat to RQM+CPL, because we have noted that observed events \emph{are} absolute within RQM+CPL: when Chidi performs a measurement on a system $S$, then there is a fact about which outcome he sees, and this fact is `absolute' in the sense that any other observers can in principle learn this fact by means of an appropriate physical interaction. Indeed, \cite{doi:10.1086/732830} has explicitly discussed this issue, arguing that `tame' relationalism does not   solve the problem raised by the EWF scenarios; only `feral' relativism does so. Here `tame' relationalism is roughly equivalent to to what I have called dynamical relationalism, with the proviso that measurement outcomes belong to the non-relativized occurrent conditions; whereas `feral' relationalism refers to a radical form of fact-based relationalism in which all physical facts are relativized. 

Now,  it  is true that `tame' or `dynamical' relationalism does not solve the problem raised by the EWF scenarios, if we assume that we are dealing with a version of  tame or dynamical relationalism which obeys Cross-Perspective Links.
But we do not have to assume that! It is certainly possible to imagine a dynamically relational theory in which there exist absolute facts about what Chidi observes in his measurement, but there is no systematic connection between Chidi's (absolute) measurement outcome $C$ and the (absolute) value $A$ that Alice hears when she asks Chidi about his outcome. And this version of tame relationalism \emph{would} solve the problem raised by the EWF, since without CPL we cannot assert an equality $A = C$ between the value learned by Alice and the value seen by Chidi, so the derivation of the contradictory predictions does not go through.  

And meanwhile, the way in which feral relationalism solves the problem raised by the EWF is also by rejecting CPL. That is, in the context of feral relationalism the  weaker RCPL postulate described in section \ref{orthodoxCPL} may hold within every individual perspective, but the absolute version cannot be upheld, since there is no fact of the matter about the `actual' result of Chidi's measurement. So in this context, the equality $A = C$ does not hold and indeed does not even make sense, and therefore the derivation of the contradictory predictions cannot go through. So ultimately, both approaches ultimately come down to rejecting the absolute version of CPL. Therefore I think the tame/feral distinction is a red herring here: what really matters here is the absolute version of the CPL postulate, and it is possible to deny that postulate within either  tame or feral relationalism\footnote{The only substantive difference here is that in feral relationalism we are \emph{obliged} to deny CPL, whereas in tame relationalism we have a choice about it. Since the denial of CPL leads to significant epistemic problems, this does not look like  an advantage for feral relationalism.}.

Of course, for the reasons  described in section \ref{intersubjectivity}, I actually think that regardless of whether one prefers tame or feral relationalism, it is a bad idea to deny the absolute version of CPL in general: this move would severely undermine the whole epistemology of science. In particular, as emphasized in section \ref{intersubjectivity},  the weaker RCPL postulate is not adequate to make sense of the epistemology: in order to empirically confirm a relational version of quantum mechanics it must be possible to learn about facts for which the relativization terminates in perspectives other than one's own, so we need the absolute version, i.e. the version which does support the equality $A = C$. Thus it is implausible that getting rid of CPL \emph{in general} can possibly be the right solution to the EWF paradox.

With that said, there might possibly be a viable route where we maintain that the (absolute) CPL postulate does hold in general, but posit that it fails in the particular circumstances relevant to the EWF experiment. As long as we have clear principled reasons to explain why such a thing would occur in these circumstances but not in most ordinary instances of scientific communication, the epistemology of science more generally would seem to survive  unscathed. But a view of this kind would not be anything like feral relationalism or Orthodox RQM, since it would maintain the existence of an `absolute' shared reality except in some very special cases. 

Conversely, if we insist that the absolute version of CPL  holds  in the context of the EWF scenarios, then the EWF scenarios are not really an argument for \emph{any} kind of relationalism: regardless of whether or not we adopt some form of relationality, if we want to maintain absolute CPL together with the first-person universality of quantum mechanics, we will either have to  either reject the Single-World part of AOE, leading to an Everettian picture, or we will have to reject one of the other assumptions, i.e. Locality or No-Superdeterminism. 

So what significance \emph{do} the EWF theorems have for relationalism? Well, their most important consequence is that    any  relational view which is not a many-worlds view and which upholds first-person universality while also maintaining (absolute) CPL will necessarily have to violate either Locality or No-Superdeterminism or both. And indeed, \cite{https://doi.org/10.48550/arxiv.2203.13342} explicitly acknowledge that RQM+CPL should probably be understood as both non-local and retrocausal, so a solution to the dilemma posed by the EWF theorems is already baked into that model.

\section{Conclusion}

My goal in this article has been to  elaborate on the alternative version of RQM originally suggested by \cite{https://doi.org/10.48550/arxiv.2203.13342}. I have suggested that this proposal can be motivated as a way of combining the commitment to \emph{inherent} relationality on which views like RQM are based with the dynamical form of relationalism we see in views like Everett and de Broglie-Bohm, and that there are good reasons to think that a view of this kind is worth pursuing. 

I have also have argued that all else being equal, we should prefer RQM+CPL over Orthodox RQM, since it offers a more coherent account of various aspects of epistemology such as intersubjectivity, our ownership of our own experiences, and our knowledge of our own condition. I have also demonstrated that worries concerning the regress of relativization and EWF theorems can be dealt with.   Thus I suggest that RQM+CPL, and possibly other inherent-dynamical approaches, offers an appealing alternative to other relational approaches, and should be of interest to anyone who feels the force of the  intuitions driving relational approaches to quantum mechanics.

\section{Acknowledgements:} Thanks to Carlo Rovelli, Timotheus Riedel and Eric Cavalcanti for some helpful conversations on subjects related to this paper.

\end{document}